\def\ltsim{\raise 2pt \hbox {$<$} \kern-1.1em \lower 4pt \hbox {$\sim$}}
\def\ltapprox{\raise 2pt \hbox {$<$} \kern-1.1em \lower 5pt \hbox {$\approx$}}
\def\gtsim{\raise 2pt \hbox {$>$} \kern-1.1em \lower 4pt \hbox {$\sim$}}
\def\gtapprox{\raise 2pt \hbox {$>$} \kern-1.1em \lower 5pt \hbox {$\approx$}}
\documentstyle[11pt]{elsart}    
\def\ltsim{\raise 2pt \hbox {$<$} \kern-1.1em \lower 4pt \hbox {$\sim$}}
\def\ltapprox{\raise 2pt \hbox {$<$} \kern-1.1em \lower 5pt \hbox {$\approx$}}
\def\gtsim{\raise 2pt \hbox {$>$} \kern-1.1em \lower 4pt \hbox {$\sim$}}
\def\gtapprox{\raise 2pt \hbox {$>$} \kern-1.1em \lower 5pt \hbox {$\approx$}}

\def\skuno{\vskip 20pt}

\begin{document}
\begin{frontmatter}
\title{Particle injection and reacceleration in clusters of galaxies  
       and the EUV excess: the case of Coma}
\skuno
\author[io]{G. Brunetti\thanksref{b}},
\author[io]{G. Setti\thanksref{s}},
\author[lu]{L. Feretti\thanksref{f}},
\author[gg]{G. Giovannini\thanksref{g}}

\address[io]{
       Dipartimento di Astronomia, Universit\'a di Bologna, 
       via Ranzani 1,
       I--40127 Bologna, Italy.
       Istituto di Radioastronomia del CNR, via Gobetti 101,
       I--40129 Bologna, Italy.}
\address[lu]{
       Istituto di Radioastronomia del CNR, via Gobetti 101,
       I--40129 Bologna, Italy.}
\address[gg]{
       Dipartimento di Fisica, Universit\'a 
       di Bologna, via Berti--Pichat 6/2, 
       I--40127 Bologna, Italy.
       Istituto di Radioastronomia del CNR, via Gobetti 101,
       I--40129 Bologna, Italy.}

\thanks[b]{gbrunetti@astbo1.bo.cnr.it}
\thanks[s]{setti@astbo1.bo.cnr.it}
\thanks[f]{lferetti@ira.bo.cnr.it}
\thanks[g]{ggiovann@ira.bo.cnr.it}

\begin{abstract}

We calculate the energy distribution 
of the relativistic particles injected in the
ICM during a phase of reacceleration of the relativistic
particles in the cluster volume.
We apply our results to the case of the 
Coma cluster in which recent merging activity
and the presence of the radio halo 
may suggest
that reacceleration processes are efficient.
We find that the electron population 
injected in the
central part of the cluster by the head--tail
radio galaxy NGC 4869 may account
for a large fraction, if not all, of the
detected EUV excess via inverse Compton 
scattering of the cosmic microwave 
background (CMB) photons.

If radio haloes are powered by
reacceleration mechanisms active in the
cluster volume, moderate non--thermal
EUV excesses (of order of 
$\sim 1-5 \cdot 10^{42}$erg s$^{-1}$)
should be a common feature of clusters containing powerful 
head--tail radio galaxies and/or AGNs.


\end{abstract}


\end{frontmatter}

\section{Introduction}

Observations of a few galaxy clusters 
with the {\it Extreme Ultraviolet Explorer} 
have revealed extreme ultraviolet emission  
(EUV) in excess of that expected by 
extrapolating downward the thermal 
X--ray spectrum 
(Lieu et al. 1996a,b; Bowyer, Lampton
and Lieu 1996; Mittaz, Lieu and 
Lockman 1998).
The EUV excess 
has been claimed 
by these authors as a new feature of clusters
of galaxies with important physical implications.

The suggestion that spurious EUV excesses may  
derive by underestimating
the line--of--sight Galactic absorption
(Arabadjis \& Bregman 1999)
has been excluded 
by further EUV investigations
(Bowyer, Berg\"ofer and Korpela 1999).
These authors have also pointed out that the measured EUVs  
are strongly 
influenced by the variation of the telescope
sensitivity over the field of view.
By making use of the appropriate corrections,
EUV extended excess was confirmed in the case of
Coma (Bowyer et al. 1999) and Virgo 
(Bergh\"{o}fer, Bowyer and Korpela 2000), whereas  
no evidence of extended EUV excess was found
in A1795 and A2199 (Bowyer et al. 1999).
The presence of EUV excess in the latter two
clusters is, however, still debated
(Bonamente, Lieu and Mittaz 2000; Lieu, Bonamente and 
Mittaz 2000).

The EUV excess may be interpreted as due to 
a relatively cool 
($\sim 10^6$ K) emitting gas (e.g. Lieu et al.1996a), 
but, at these temperatures the gas
cooling is particularly efficient, 
so that a non--thermal interpretation 
is usually favoured with respect to  
the thermal scenario 
(Hwang 1997; Ensslin \& Biermann 1998;
Bowyer \& Bergh\"ofer 1998; Sarazin \& Lieu; Lieu et al. 1999).
Indeed, synchrotron emission in the radio band and  
inverse Compton (IC) emission in the EUV and hard X--rays 
are expected from clusters of galaxies
in the
framework of continuous injection of   
relativistic electrons (Sarazin 1999; V\"{o}lk 
\& Atoyan 1999) and of a  
{\it two phase} model invoking reacceleration of 
previously injected relic particles 
(Brunetti et al. 2000).

Due to the large amount of data the Coma 
cluster represents the best case to
compare model predictions with observations.
Since the EUV profile  
is narrower than the radio one, it cannot be
accounted for by IC emission of the low energy
halo electrons with the CMB photons 
(Bowyer \& Bergh\"{o}fer 1998).
Thus, alternative scenarios have been
explored.

Ensslin et al. (1999) have calculated 
that anisotropic IC scattering
of the optical photons (from the central galaxies
in the cluster) by an assumed ad hoc
very dense population 
of mildly relativistic electrons (with a  
cut--off in the energy distribution at a few MeV) 
can, in principle, account for the EUV properties.

Atoyan \& V\"{o}lk (2000) have argued that the 
EUV can be accounted for by IC scattering of the CMB
photons from a relic electron
population with a radiative cut--off at 
$\gamma \geq 500$, 
injected in the ICM 
at the epoch of starburst activity
and recently slightly re--energized.

Sarazin (1999; 2000) has pointed out that clusters of
galaxies can contain a large number of relic
relativistic electrons (with $\gamma < 300$) injected
in the past which can account for the EUV excess
via IC scattering of CMB photons.

In a recent paper Brunetti et al.(2000) have shown that 
the radio spectral steepening observed in the case of Coma C
(Giovannini et al. 1993; Deiss et al. 1997) 
may imply the presence of diffuse
reacceleration in the cluster volume and the spectrum
of a relic population would be stretched toward higher 
energies thus reducing the emitted EUV flux.

Following the original proposal of Brunetti et al.(1999), 
in this paper we investigate the possibility that 
a substantial fraction, 
if not all, of the EUV excess emission from the Coma
cluster can be accounted for by the IC scattering
of the CMB photons due to relativistic
electrons recently injected in the central part of
the cluster by AGN activity and reaccelerated by 
moderate turbulence in the ICM.

In Sections 2.1 and 2.2 we illustrate the proposed scenario 
and in Section 2.3 we discuss the possibility that NGC 4869
may supply the required number of relativistic 
particles; a general discussion is given in Section 3.
In Appendix A we calculate the 
time evolution of the energy distribution 
of relativistic electrons continuously injected and
reaccelerated in the ICM, while 
relevant formulae used to calculate the synchrotron
and IC emitted spectra are given in the Appendix B.

$H_0 = 50$ km s$^{-1}$ Mpc$^{-1}$ 
is assumed throughout.

\section{Modelling the EUV excess in the Coma cluster}

\subsection{The proposed scenario}

We assume the scenario of the {\it two phase} 
model (Brunetti et al. 2000) in which relic 
relativistic electrons, continuously injected in the 
cluster volume in the past, have been recently 
reaccelerated by diffuse shocks/turbulence probably 
induced by a cluster merger.
These electrons 
(the main electron population, {\it MeP})  
are responsible for both the 
diffuse radio emission and at least for a large fraction 
of the hard X--rays emitted by Coma via the IC
scattering of the CMB photons.
The resulting IC spectrum is however flat and falls well
below the observed EUV flux. 

As already stated in the Introduction, 
the brightness profile of the EUV excess is
much narrower than the radio profile 
(Bowyer et al. 1999 and references therein)
so that, an IC origin of the
EUV excess should involve an additional
electron population ({\it AeP}).
The requirement is that 
the synchrotron radio contribution from
the {\it AeP} should be 
considerably smaller than that from the
{\it MeP} ($\leq 20-30\%$), i.e. 
the energy distribution of the {\it AeP}
should be considerably steeper than 
that of the {\it MeP}.

In order to preserve a relatively steep energy
distribution during a reacceleration phase
it is necessary that the injection of the
{\it AeP} has started 
recently, i.e. not more than $\sim$3 times the 
reacceleration time.
The problem, in this case, is that 
an efficient injector of
relativistic electrons is required to match 
the energetics necessary to account for the
measured EUV.
Although the main consequences of this scenario 
do not depend on the details of the injection,
we examine and propose that it can 
be identified with the recent activity of
relatively powerful radio loud AGNs, such as the head 
tail radio source NGC 4869.

In order to test this hypothesis, in the next Section
we model the time--evolution of
the {\it AeP} energy distribution injected 
in a turbulent ICM.

\subsection{The model}

We consider a simplified
model based on the following assumptions:

a) the {\it AeP} is uniformly injected in the cluster core
(spherical geometry) at a constant rate for a time 
interval $\Delta t_{inj}$ and with an energy spectrum, typical
of the radio galaxies, of the form:

\begin{equation}
Q_{inj}(\gamma)= K_e \gamma^{-\delta}
\left( 1- {{\gamma}\over{\gamma_{b}^{rg} }} \right)^{\delta-2}
\label{inj_rg}
\end{equation}

\noindent
where $\gamma$ is the electron Lorentz factor, $\gamma_b^{rg}$ the 
break energy and $\delta > 2$;

b) following Sarazin (1999) the electrons lose their energy by 
synchrotron and Compton losses, 
the last being dominant because of the relative weakness of
the cluster magnetic fields, and at low energies by Coulomb losses,
whereas they are continuously reaccelerated by systematic Fermi
processes ($d \gamma / dt \propto \gamma$).

The evolution of the reaccelerated relativistic 
particles changes depending on wether or not 
$\gamma_b^{rg}$ is larger 
than the asymptotic break energy $\gamma_b(\infty)$ 
resulting from the
balance between energy losses and gains in the ICM.
The time evolution of the {\it AeP} energy spectra are represented 
in Fig.1 and Fig.2 for these two relevant cases and for values of the
parameters representative of the astrophysical problem
being discussed (the detailed derivation of the formulae 
is given in the Appendix A).
Fig.1 illustrates the case of $\gamma_b^{rg} > 
\gamma_b(\infty)$.
With increasing $\Delta t_{inj}$ an increasing number of 
electrons is accumulated below $\gamma_b(\infty)$, 
the spectrum flattens 
at lower energies due to combined 
reacceleration and Coulomb losses, while
at higher energies (but $<< \gamma_b^{rg}$) 
it is $\propto \gamma^{-\delta+1}$
as in the standard case of continuous injection.

Fig.2 shows the case in which
$\gamma_b^{rg} < \gamma_b(\infty)$; again the electron 
spectrum flattens with increasing $\Delta t_{inj}$, but a
sharp cut--off is maintained at 
$\gamma \leq \gamma_b(\infty)$.

We have then applied standard formulae (Appendix B) to
compute the IC and synchrotron emissions from a spherical
volume of 15 arcmin radius (about the extension of
EUV excess)
by adopting the physical parameters 
consistent with the two phase model of Brunetti et al.(2000)
for Coma.
These are: a magnetic field of average strength $0.5-0.6
\mu G$ (within $\sim 15$ arcmin), 
an acceleration parameter $\chi = 3.5 \cdot 10^{-16}$s$^{-1}$
and a Coulomb loss coefficient $\xi = 1.3 \cdot 10^{-15}$s$^{-1}$.
With these values $\gamma_b(\infty) \simeq 1.8 \cdot 10^4$ 
and electrons of this energy would typically emit synchrotron
radiation at a frequency $\nu \sim 100 $MHz.
In the framework of the present model there are then two 
consequences which follow from the
requirement that the predicted radio emission should have a
steep spectrum in order not to significantly contribute to
the 327--1400 MHz radio spectrum of the
Coma cluster: first, the injection period $\Delta t_{inj}$
cannot be larger than $\sim 0.3$ Gyr and, second, 
$\gamma_b^{rg}$ must be significantly smaller 
than $\gamma_b(\infty)$ (Fig. 1 and 2).

By assuming $\delta = 2.6$, typical of radio galaxies, and
$\gamma_b^{rg}=1000$ we have normalized the electron spectrum by 
requiring that the IC scattering with the CMB photons matches the 
observed EUV excess and computed the IC and synchrotron 
emissivities  as shown in Fig. 3 and 4.

It is seen that the model accounts for the EUV flux without
introducing any significant contribution to the IC X--ray
flux derived by Brunetti et al.(2000) model, whose
parameters have been adopted here, but
the predicted synchrotron emission at low radio
frequencies may be significant and may
account for the apparent upturn of the radio spectrum
indicated by the measurements at low frequencies.
If the 74 MHz flux is significantly contributed
by the additional electron population 
(the case $\Delta t_{inj} =0.3$ Gyr in Fig.4)
the 74--327 MHz synchrotron spectrum within $\sim$15 arcmin 
should be steeper ($\alpha \geq 1$) than that between 
327--1400 MHz; detailed VLA observations at 74 MHz
would help in testing this.

It should also be noticed that in the model the 
values of $\gamma_b^{rg}$ and $\Delta t_{inj}$ are
(roughly) inversely correlated: for instance, the 
adoption of a somewhat higher value for
$\gamma_b^{rg}$, e.g. 2000, would still be consistent with 
the low
frequency radio data by simply reverting to a smaller value
of $\Delta t_{inj}$.

\begin{figure}
\includegraphics{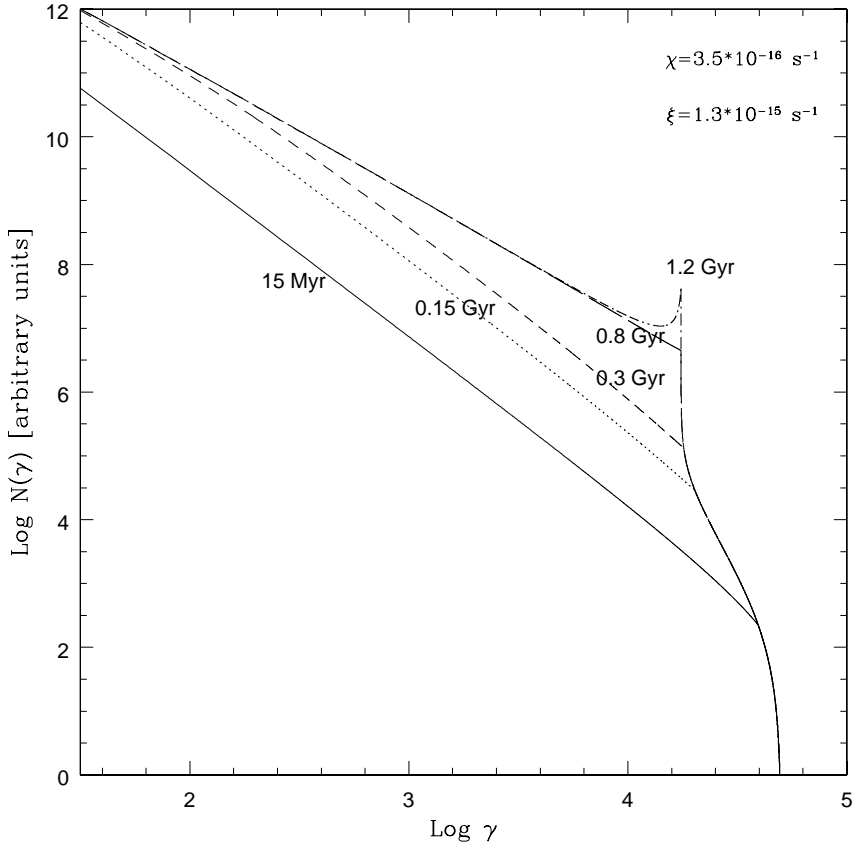}
\vspace{12 cm}
\caption{The energy distribution 
of electrons injected
in the ICM and reaccelerated 
is reported in arbitrary units.
For each curve 
the injection/reacceleration period $\Delta t_{inj}$
is given in the panel.
The assumed coefficients of reacceleration 
and of Coulomb losses are 
$\chi=3.5\cdot 10^{-16}$s$^{-1}$
and 
$\xi=1.3\cdot 10^{-15}$s$^{-1}$, respectively; the other 
parameters are $B=0.5 \mu$G, $\gamma_b^{rg}=50000$ 
and $\delta=2.6$.} 
\end{figure}

\begin{figure}
\includegraphics{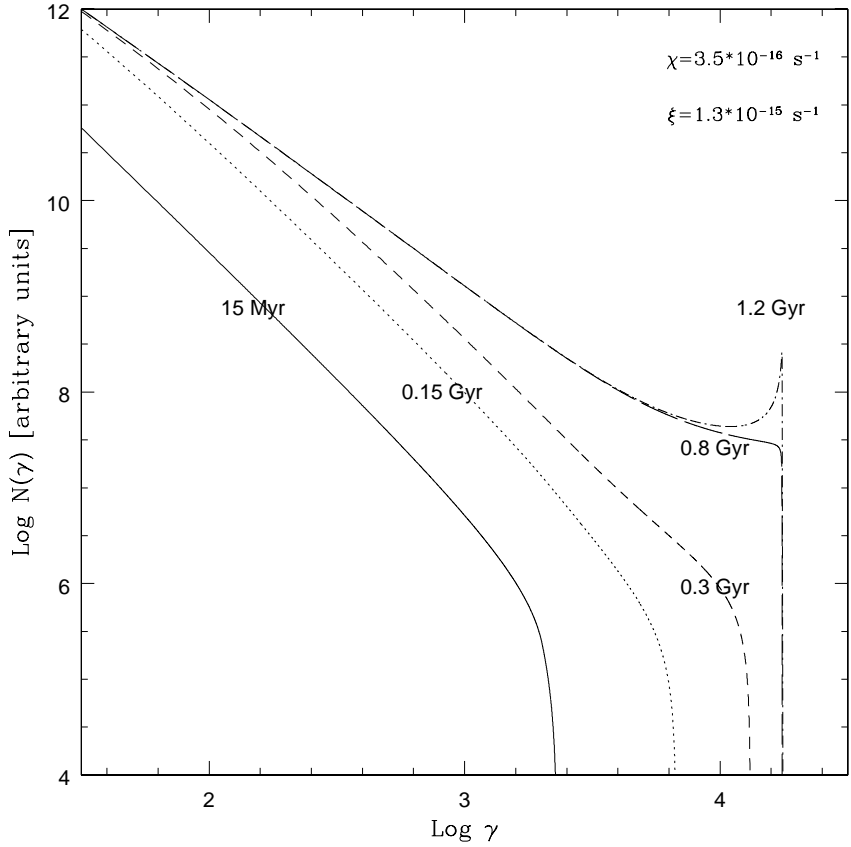}
\vspace{12 cm}
\caption{The calculated 
electron energy distribution 
of electrons continuously reaccelerated and injected
in the ICM is reported in arbitrary units.
In this case the break energy of the injected 
population, $\gamma_b^{rg}=2000$, is
smaller than $\gamma_b(\infty)$; 
the other parameters being the same of Fig.1.} 
\end{figure}

\subsection{NGC 4869 and the EUV excess in Coma}
 
In this Section we compare the energetics of the electron 
population required 
by the model to match the EUV excess via IC scattering
of CMB photons by 
the electron population continuously released in the ICM 
by the AGNs in the Coma cluster.
NGC 4869, the head tail radio source, is presently the 
most powerful radio galaxy in the core of the Coma cluster.
The question is whether or not it may supply a sufficient
fraction of the energetic electrons required to explain
the EUV excess according to the model described
in the preceding Section 2.1.

Dallacasa et al.(1989) have reported three 
frequency (327, 1465, 4885 MHz) 
spectra of NGC 4869 up to a distance of 
$\sim 180$ arcsec from the nucleus and the 327--1465 
spectral index up to $\sim 280$ arcsec.

Feretti et al.(1990) have fitted the synchrotron
spectrum of NGC 4869 at different distances from
the radio nucleus up to 180 arcsec
finding that the break frequency 
ranges from $>$ 50 GHz 
close to the nucleus (where the slope of the emitted 
spectrum is $0.8$, i.e. $\delta = 2.6$)
to $<$ 1 GHz at larger distances.
They interpreted the 
systematic decrease of the break frequency with distance 
from the nucleus as due to the ageing of the emitting
electrons.

Since we are particularly interested in constraining
the electron population at the end of
the tail, where electrons are mixed with the ICM,
we have fitted the spectral indices of the external
part of the tail reported by
Dallacasa et al.(1989).
We have assumed $\delta = 2.6$ and a KP model
(Kardashev 1962; Pacholczyk 1970)
which corresponds to a non re--isotropization of the
electron momenta due to the dominance of the synchrotron
losses in the tail volume
(the SYNAGE fitting software package
developed by Murgia \& Fanti 1996 has been used).
At distances $> 220$ arcsec
the break frequency is not well
determined ($\nu_b$\ltsim$200$ MHz)
since the 327--1465 MHz spectral index
approaches the KP--asymptotic value $\sim 2$.
By considering the spectral fits and
by assuming the particle age to be
proportional to the distance from the nucleus
(in agreement with Feretti et al. (1990) findings)
a break energy
$\gamma_b$\ltsim$4000/ \sqrt{ B_{\mu G} }$
is estimated in the oldest detectable parts
of the radio tail ($\sim 280$ arcsec).
It should be stressed that this break energy is only 
an upper limit to $\gamma_b^{rg}$ in 
Eq.(\ref{inj_rg}) since the detected radio flux is 
emitted
by plasma still confined in the radio tail and
not yet well mixed with the ICM.
We conclude that $\gamma_b^{rg} \sim 1000$
(given a magnetic field of the radio galaxy
$B \sim 10 \mu$G; see below), 
adopted in the model of the preceding Section 2.2, 
is appropriate to the case of NGC 4869.

\begin{figure}
\includegraphics{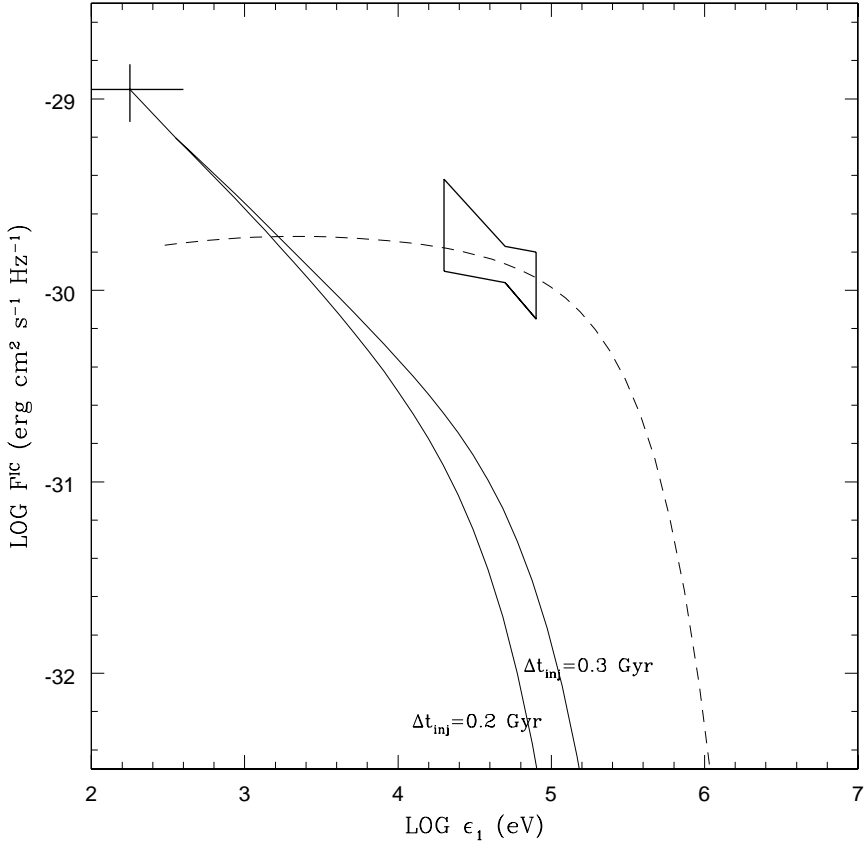}
\vspace{12 cm}
\caption{The calculated IC spectrum  
from the {\it AeP}
injected in the ICM medium is reported for
$\Delta t_{inj}$= 0.2 and 0.3 Gyr (solid lines).
The curves are normalized to the flux of the EUV excess.
In the calculation we have assumed 
$\gamma_b^{rg}=1000$, $\delta=2.6$ and
$\chi=3.5 \cdot 10^{-16}$s$^{-1}$.
The expected IC contribution from the 
{\it MeP} in the
framework of the {\it two phase} model 
(dashed line) is taken by Brunetti et al.(2000).} 
\end{figure}

\begin{figure}
\includegraphics{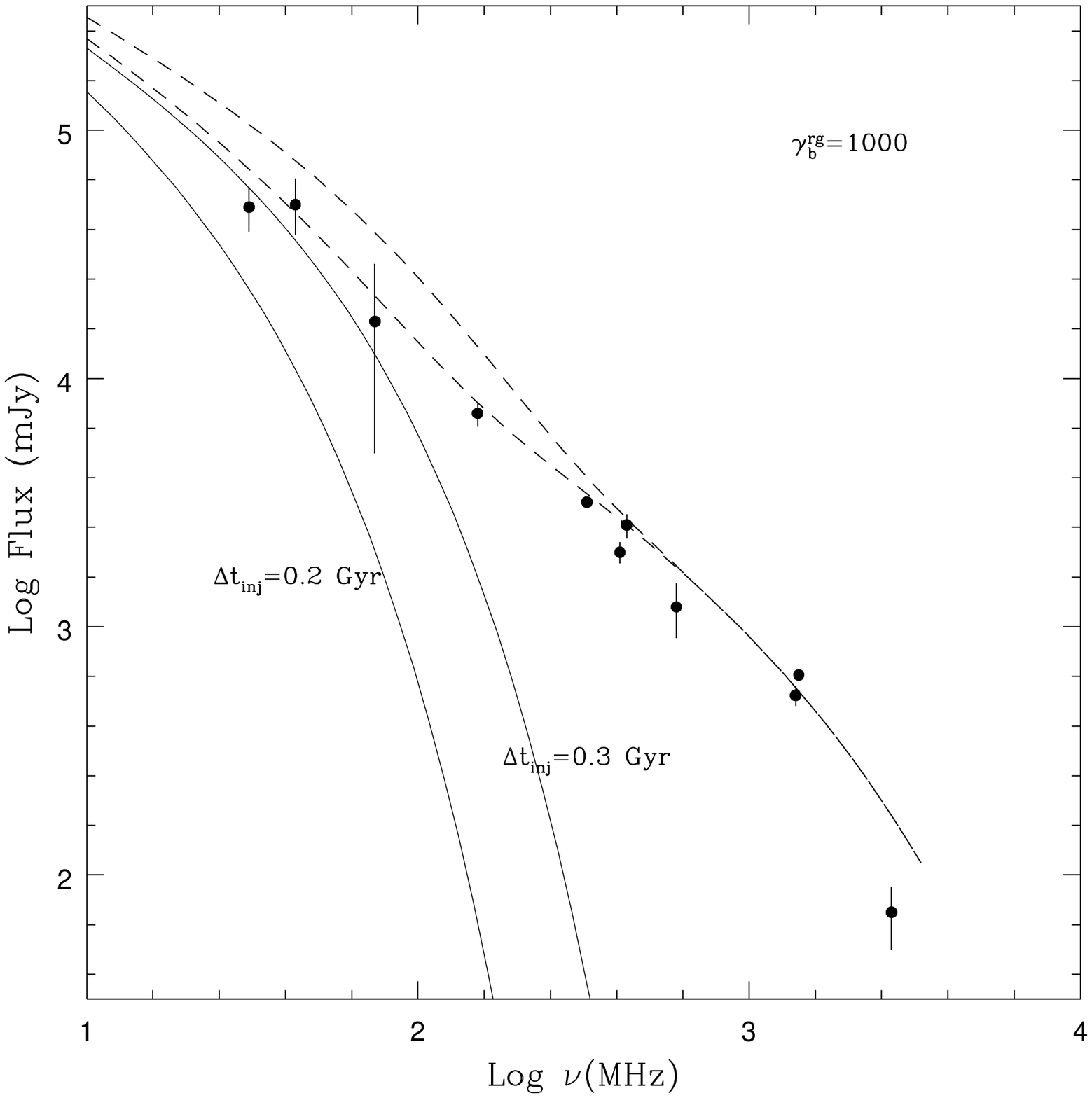}
\vspace{12 cm}
\caption{The calculated synchrotron spectra 
from the {\it AeP}  
injected in the ICM (solid lines) are reported for
the IC models represented in Fig.3.
A volume averaged $B=0.6 \mu G$ is assumed.
The dashed lines represent the total
synchrotron spectrum from Coma C, i.e. the
sum of the contributions from the {\it AeP} 
and from the reaccelerated relic particles of the 
{\it two phase} model (taken from Brunetti et 
al. 2000).} 
\end{figure}

Under the simple assumption of an injection rate  
($Q_{inj} = \int Q_{inj}(\gamma) d\gamma$) 
constant with time, 
the total electron number injected in the cluster 
volume is $Q_{inj} \Delta t_{inj}$.
In order to test if 
NGC 4869 can account for the 
total number of electrons 
necessary to match the EUVs, we have calculated
$Q_{inj}$ between 30 and 60 arcsec
from the nucleus of the radio galaxy 
under minimum energy conditions.
Since the EUV flux is due to IC by the 
electrons of the {\it AeP} with Lorentz factor 
$\gamma \sim 300$, the classical equipartition
formulae are not sufficient and we should include 
low energy electrons in the computation of
the minimum energy parameters  
(Brunetti et al. 1997).
By assuming a low energy cut--off in the injected
electron spectrum $\gamma_{min} =50$ 
(below which Coulomb losses in the tail  
would flatten the electron distribution), a ratio = 1 between 
proton and electron energies and a filling factor = 1, 
we obtain an 
equipartition magnetic field 
in the radio galaxy $B_{eq} \sim 14.5 \mu G$
at a projected distance between 30 and 60 arcsec (20--40 kpc) 
from the nucleus.
The corresponding radiative age $\sim 4.5 \cdot 10^6$yr, combined
with the synchrotron brightness in the same region
(corrected for the steepening of the spectrum), gives  
$Q_{inj} \sim 6.5 \cdot 10^{54}$el yr$^{-1}$.

By normalizing Eq.1 to this value, we find that
the IC contribution to the observed EUV flux 
from the {\it AeP} injected in the ICM 
is $\sim 30 \%$ and $\sim 40 \%$ for 
$\Delta t_{inj} =$ 0.2 and 0.3 Gyr, respectively. 
Although this time could be long if compared
with the typical age of the radio galaxies, 
we point out that it is    
similar to the ages estimated in the case of some 
head tail radio galaxies (e.g. IC 310 and 3C 129, Feretti 
et al. 1998) and in the case of giant radio galaxies 
(e.g. B 0313+683, Schoenmakers et al. 1998; 
B 0319-454, Saripalli, Subrahmanyan, Hunstead 1994).

There is no firm evidence so far of a
perfect energy equipartition between magnetic
fields and relativistic particles in the radio 
galaxies, thus we have also investigated 
the consequences of a moderate departure
from equipartition.
By assuming a magnetic field strength lower than 
the equipartition value both the number density  
of the relativistic electrons and the 
radiative age increase, thus $Q_{inj}$ does not 
critically depend on the departure from the  
equipartition condition.
Specifically we obtain :

\begin{equation}
{{Q_{inj}(B)}\over{Q_{inj}(B_{eq}) }}
=
\left( {{B_{eq} }\over{ B}} \right)^{ {{\delta+2}\over 2} }
{{
\left[ 1 + 
\left(
{{B}\over{B_{IC}} } 
\right)^2 
\right]
}\over{
\left[ 1 + 
\left(
{{B_{eq}}\over{B_{IC}} } 
\right)^2 
\right]
}}
\label{re}
\end{equation}

\noindent
where $B_{IC} \sim 3 \mu$G is the equivalent field of the CMB.
Eq.(\ref{re}) is reported in Fig.5 : a net increase 
of $Q_{inj}$
requires a magnetic field $B < 0.2$ $B_{eq}$.
\begin{figure}
\includegraphics{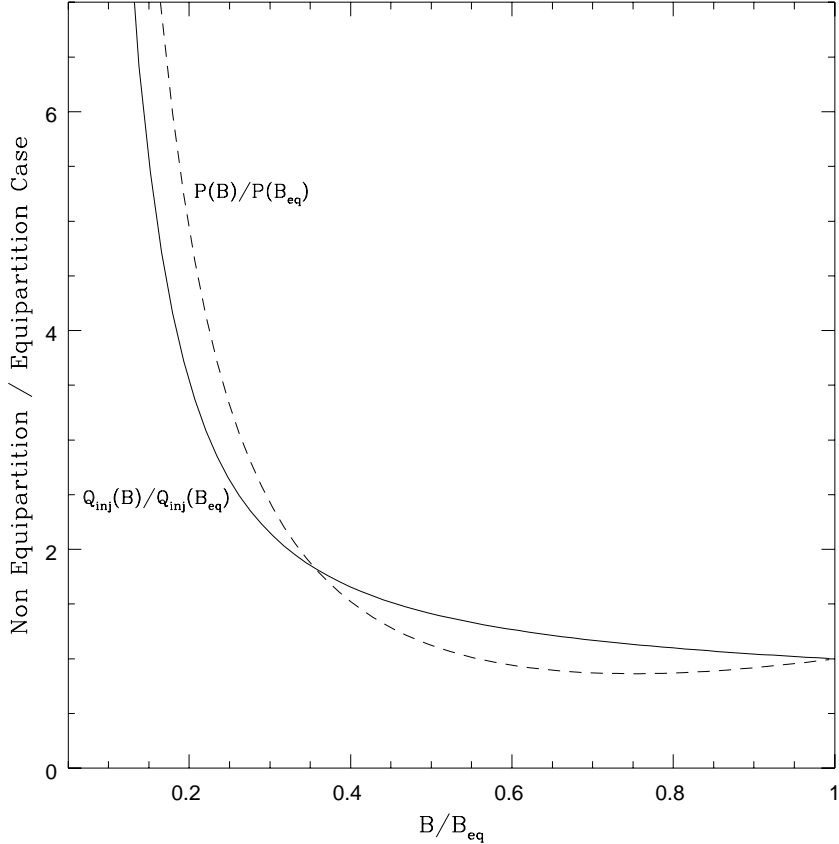}
\vspace{12 cm}
\caption{
The particle injection rate (solid line)
and the internal pressure (dashed line)
normalized to the corresponding values
under equipartition conditions are reported for 
$\gamma_{min}=50$ and $\delta=2.6$ ($B< B_{eq}$).} 
\end{figure}
In order to match the EUV flux a magnetic field
strength $B \sim 1/3-1/4 B_{eq}$ is required. 

Implications on the pressure equilibrium
between the tailed radio source and the ICM can
finally be obtained.
By making use of standard equipartition equations, 
Feretti et al. (1990) have shown that NGC 4869 is
in apparent imbalance with the external thermal 
pressure by a factor $\sim 10-15$ in the region
between 30--60 arcsec and by a larger factor
with increasing distance from the nucleus;
this imbalance is partly reduced by
considering the contribution of the low energy
electrons.
We have calculated the expected pressure 
inside the radio volume under our assumptions.
The equipartition pressure at 30--60 arcsec from the 
nucleus is 
$\simeq 10^{-11}$dyne cm$^{-2}$,
while it is  
$\simeq 4 \cdot 10^{-12}$dyne cm$^{-2}$
with standard formulae.
 
Out of equipartition (for an ordered magnetic field)
the pressure
can be obtained from Brunetti et al.
(1997) equations; one has :

\begin{equation}
{{P(B)}\over{P(B_{eq})}}=
{{ 3\delta +3}\over{3\delta +7}}
\left( {{B}\over{ B_{eq} }} \right)^2
\left(
1+
{1\over 3}
{4\over{\delta+1}}
\left( {{B_{eq}}\over{B}} \right)^{ 
{{\delta +5 }\over 2} }
\right)
\label{pressureuv}
\end{equation}

\noindent
which is reported in Fig.5.
Since under our hypothesis
the particles contribute to the pressure
for only $1/3$ of their energy, 
a moderate departure from equipartition causes a
slight decrease of the pressure, which however, rapidly 
increases for larger departures.  

By assuming the magnetic field in NGC 4869  
required to match the EUV excess via IC scattering, the internal pressure
increases up to $\sim 2-2.5 \cdot 10^{-11}$dyne cm$^{-2}$ 
which is rather close (but a factor $\sim 2-2.5$ below) 
to the thermal pressure.

\section{Discussion and Conclusions}

It has been previously shown (Brunetti et al 2000) that the radio 
(synchrotron) and hard X-rays (IC on the CMB photons) properties of
Coma C may be explained by a two phase model in which diffuse 
relic electrons have been efficiently reaccelerated throughout the
cluster volume for a period of $\sim$1 Gyr by the turbulence and 
associated
Fermi type mechanisms excited in the ICM by a recent merger
(main electron population, {\it MeP}). 

By adopting this model we have shown here that a steady injection in the
cluster core of an additional 
fresh population of relativistic 
electrons (Fig.6)
may explain the EUV excess, recently determined by Bowyer et
al.(1999), as IC scattering of the CMB photons. 
Detailed computations
of the time evolution of the injected electron spectrum show that
internal consistency within the complete model requires a relatively
steep injection spectral index ($\delta \sim 2.6$), 
however typical of the
radio galaxies, and an injection starting time $\leq 0.3$ Gyr, 
in which 
case there is no significant contribution to the synchrotron emission
in the 327-1400 MHz band and to the total X-ray flux. 
Nevertheless, it 
should be pointed out that an important contribution to the emission
at low radio frequencies ($\leq 100$ MHz; Fig.4) may be present, 
so that improved measurements at long radio wavelengths may provide a 
test of the proposed interpretation of the EUV excess and, of course, 
constrain the model parameters.

The luminosity of the EUV excess is $\sim 1.5 \cdot 
10^{42}$ erg/sec. 
An IC origin
with the spectral shape of our model needs a total number
of injected electrons $\sim 4\cdot 10^{63}$. 
We have explored the possibility
that a large fraction of these electrons may have been injected 
by the powerful
head tail radio galaxy NGC 4869 orbiting in the cluster core. 
By assuming that NGC 4869 has been constantly active in the past 
0.3 Gyr (an age comparable to that of a number of giant and
head tail radio galaxies) 
we find, based on minimum energy conditions in the tail, that about
40\% of the required number of electrons may have been released to the
ICM; a full complement is achieved with a moderate departure from
the equipartition resulting in a magnetic field intensity about a 
factor $\sim 3$ weaker. 
The total energy of the electrons deposited in
the ICM is $\sim 4 \cdot 10^{59}$ erg 
which is not unreasonable for a radio galaxy.

We stress that, although the internal pressure is considerably 
increased with respect to previous estimates based on classical
equipartition formulae, the radio tail is still well confined by the
hot ICM: the pressure gap is reduced by almost an order
of magnitude, but the tail is still underpressurized by a factor 
2--3. The introduction of a filling factor and of a relativistic 
proton component may now easily bring to a pressure equilibrium.

In Fig.7 the observed EUV radial profile is compared with 
that predicted
by our model based on the simple assumption of a spherical and 
uniform distribution of the relativistic particles centered on the 
cluster centre. 
It is seen that the model provides a very good fit 
of the data at an angular distance $> 3$ arcmin from the centre 
which includes $\sim 90-95\%$ of the EUV excess. 
We note that in the 
innermost 3 arcmin the radial profile samples a region containing
the cD galaxy NGC 4874 so that an additional contribution to the
EUV flux may be expected. Furthermore, it should be stressed that
at smaller radii the azimuthal distribution of the EUV excess is
very sensitive to small scale brightness/temperature fluctuations
of the hot ICM.
If the relativistic electrons are mainly supplied by NGC 4869 then
the EUV brightness distribution should be related to some extent
with the orbital path followed by the tailed radio source. At 
present the tail's end is positioned at an angular distance 
of $\sim 6$ arcmin from
the cluster centre, so that a shift between the X-ray and EUV
maxima may be expected. 
However, most of the injected relativistic
plasma could have been stirred around the cluster core in the
available time scale by the turbulence generated by the random 
motion of the galaxies; this may have considerably smoothed out the 
distribution of the relativistic particles on the cluster core
size and thus that of the emitted EUV photons. 
Detailed 
mapping of the EUV excess would obviously be of great importance. 

We point out that our model, although specifically developed
for the Coma cluster, may have a wider application predicting the
existence of sizeable fluxes of excess EUV photons from galaxy 
clusters where large numbers of relativistic electrons are injected
in the cluster cores by powerful AGN activity, or other energetic
events such as a minor merger, during a reacceleration phase. 
However, due to the requirement of a contemporaneous presence of 
efficient Fermi type turbulence reacceleration and of an efficient
recent injection of relativistic electrons in the cluster core 
(i.e. the presence of powerful AGNs), we claim that 
in the clusters with radio haloes 
a non--thermal EUV excess due to the proposed scenario would 
be more rare than the hard X--ray excess.

Finally, if a relic population of relativistic electrons
is commonly associated with clusters of galaxies
(Sarazin 1999), 
the related EUV excess from IC scattering of CMB photons 
is expected to be more common in clusters without radio
halo than in those with radio halo where diffuse 
reacceleration
processes may stretch toward higher energies 
the electron spectrum.

\begin{figure}
\includegraphics{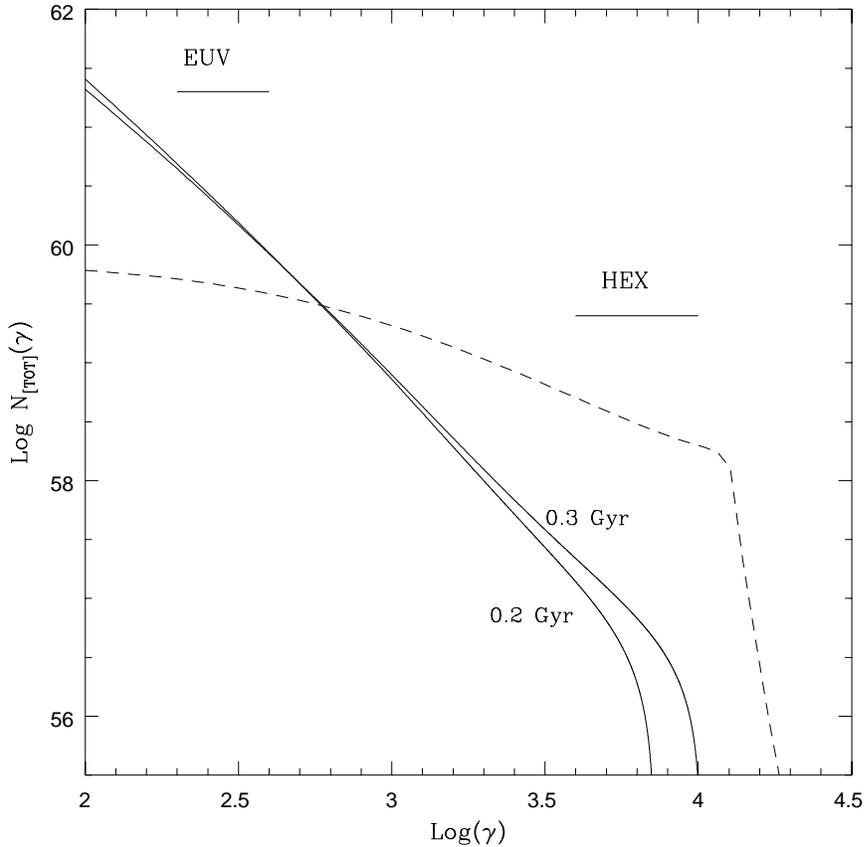}
\vspace{12 cm}
\caption{
The energy distribution integrated over the cluster
volume of the main electron population
(dashed line) and of the additional electron population
(solid lines) are shown in arbitrary units.
The typical energies of the electron emitting via IC in
the hard X--ray band (HEX) and in the EUV band are also
showed.} 
\end{figure}

\begin{figure}
\includegraphics{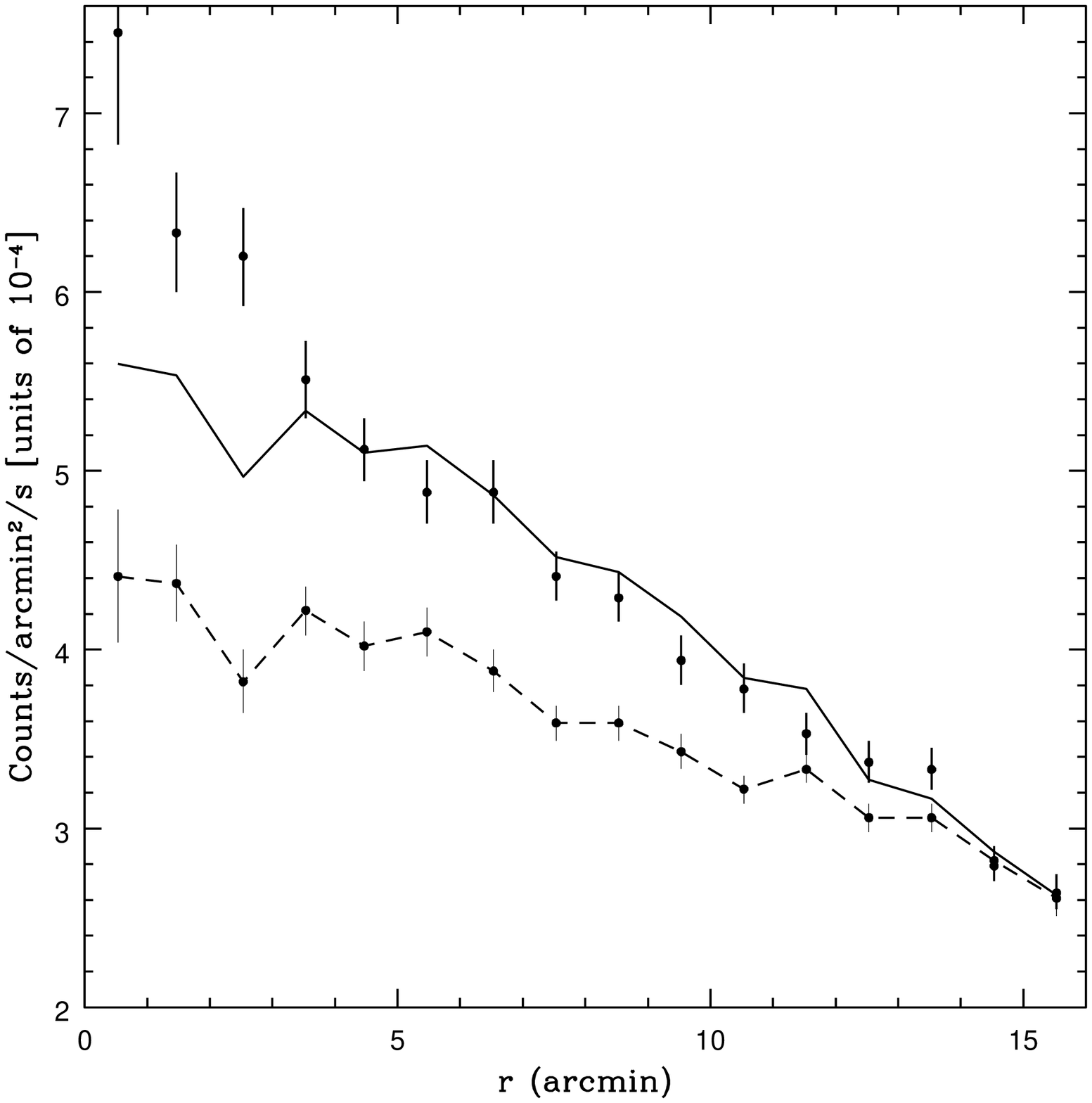}
\vspace{12 cm}
\caption{The comparison between the model calculations
and the observed EUV profile is reported.
The model EUV profile (solid line) is obtained by 
taking into account 
the instrumental background plus the expected EUV from the
X--ray plasma (dashed line).
The data for the instrumental background and EUV from
X--ray plasma  
and observed EUV excess (bold data points) are taken 
from Bowyer et al. (1999).} 
\end{figure}

\begin{ack}
We are grateful to S. Bowyer and T. Ensslin for useful 
discussions and suggestions.
This work was partly supported by the Italian Ministry for
University and Research (MURST) under grant Cofin98-02-32.
\end{ack}

\appendix{}
\section{Appendix A: Time evolution of injected and 
reaccelerated electrons}

The time evolution of the particle energy distribution
$N(\gamma,t)$ 
is obtained by solving the 
kinetic equation (e.g. Kardashev 1962):

\begin{equation}
{{\partial N(\gamma, t)}\over
{\partial t}} =
-{{\partial}\over{\partial \gamma}} 
\left[ {{d \gamma}\over{d t}} N(\gamma, t) \right]
+ Q_{inj}(\gamma, t)
\label{kinetic}
\end{equation}

\noindent
where $\gamma$ is the Lorentz factor.
We assume an injection function $Q_{inj}(\gamma)$ 
typical of 
the radio galaxies (Eq.\ref{inj_rg}). 

\noindent
Once injected, the electrons lose energy via Coulomb 
and radiation losses and are reaccelerated by
Fermi processes (here assumed to be systematic :
$d\gamma/dt = \chi \gamma$), thus, 
the evolution of the energy of the electrons is :

\begin{equation}
{{d\gamma}\over{dt}}=
\chi \gamma - \xi - \beta \gamma^2
\label{energy_evol}
\end{equation}

\noindent
where $\chi$ is the reacceleration efficiency,  
$\xi$ the coefficient of the Coulomb losses :

\begin{equation}
\xi \simeq 1.2\cdot 10^{-12} n 
\left(1 + {{ln(\gamma/n)}\over{75}} \right)
\sim 1.4 \cdot 10^{-12} n
\label{culomblosses}
\end{equation}

\noindent
and $\beta$ that of the radiative losses :

\begin{equation}
\beta(z)=
1.9 \cdot 10^{-9} \left( {2\over 3} B^2 + B_{IC}^2(z) 
\right)
\label{radiativelosses}
\end{equation}

\noindent
$B_{IC}(z)$ being 
the equivalent magnetic field strength of
the CMB.
Under these assumptions 
the time evolution of the energy of the electrons is :

\begin{equation}
\gamma(\tau)=
{{ \gamma(t_i) 
(\sqrt{q}/tanh(x) + \chi ) -2\xi }\over{
2\beta \gamma(t_i) + \sqrt{q} / tanh(x) -\chi}}
\label{gamma(t)}
\end{equation}

\noindent
with $t_i$ the time at which the injection has started,   
$q= \chi^2 -4 \xi \beta > 0$,  
$x =
{{\sqrt{q}}\over 2} \tau $ and 
$\tau$ the time interval since the injection.

\noindent
The time--evolution of the spectrum injected in the
unit time interval is
obtained by solving the kinetic equation 
with $Q_{inj}=0$. We obtain :

\begin{eqnarray}
N(\gamma,\tau)= 
K_e q 
{{ (1-tanh(x))^2}\over{ ( 2 \beta \gamma_b tanh (x) )^2 }}
\left( 1- {{\gamma}\over{\gamma_b(\tau)}} \right)^{\delta-2}
\cdot \nonumber \\
\gamma^{-\delta}
\left\{ 1 + S\left[\gamma; \gamma_b(\tau)\right] \right\}^{-\delta}
\left( 1 - {{\gamma}\over{ \gamma_b^{rg} }}
{{ 1 + S[\gamma; \gamma_b(\tau)] }
\over{ 1 - \gamma/\gamma_b(\tau)}}
\right)^{\delta-2}
\label{inj_evol}
\end{eqnarray}

\noindent
where

\begin{equation}
S[\gamma; \gamma_b(\tau)]=
{{ \xi \gamma^{-1} - \chi }\over
{\beta \gamma_b(\tau)}}
\label{s(g)}
\end{equation}

\noindent
and the break energy at time $\tau$ is:

\begin{equation}
\gamma_b(\tau)=
{1\over {2\beta}}
\left\{ {{\sqrt{q} }\over{ tanh(x)}}
+\chi \right\} \,\,
{ \buildrel \tau \rightarrow \infty  \over
\longrightarrow } \,\,
{{ \sqrt{q} + \chi}\over{2 \beta}}
\equiv
\gamma_b(\infty)
\label{gb(tau)}
\end{equation}

\begin{figure}
\includegraphics{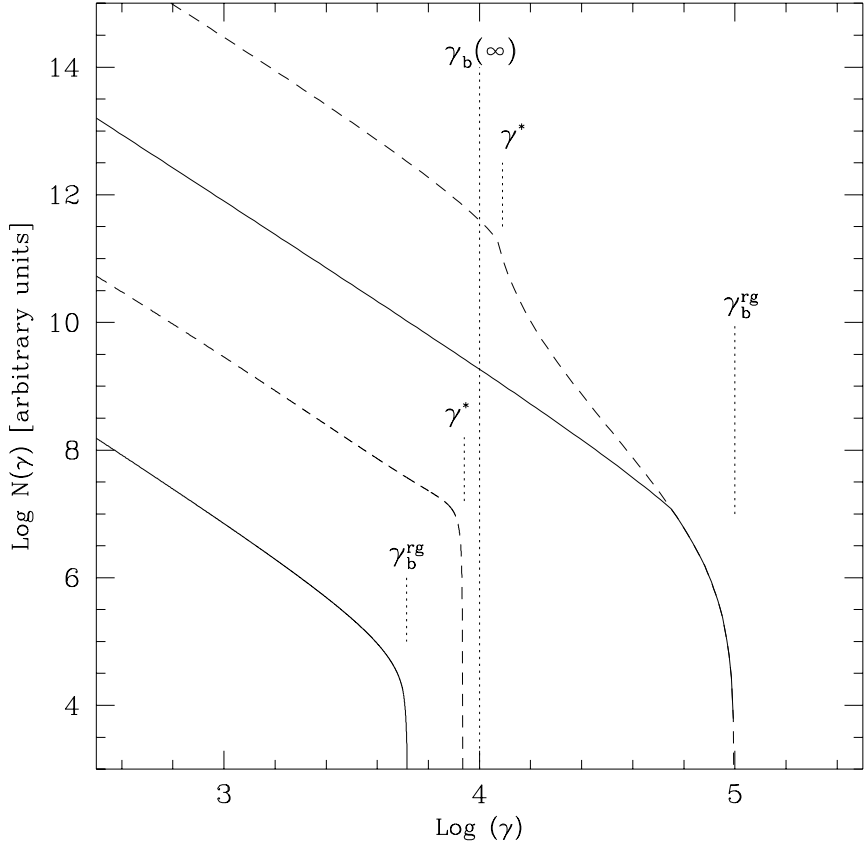}
\vspace{12 cm}
\caption{
Sketch of the electron spectrum evolution.
In the picture the asymptotic break energy is 
$\gamma_b(\infty)=10^4$. 
The injected spectra 
(solid lines) and those calculated after
an injection time $\Delta t_{inj}$ (dashed lines)
are given both in the case 
$\gamma_b^{rg} > \gamma_b(\infty)$ 
and $\gamma_b^{rg} < \gamma_b(\infty)$.
The number of electrons, in arbitrary units, 
has a different normalization
for the two cases.} 
\end{figure}

\noindent
The solution of Eq.(\ref{kinetic})
after an injection and reacceleration
period $\Delta t_{inj}$ is then obtained
by integrating Eq.(\ref{inj_evol}) over the
injection time. 
However, for a given 
Lorentz factor $\gamma$,  
only the contributions of the injected electrons
with a maximum energy larger than 
$\gamma$ should be considered.
In order to do this we introduce the 
largest energy of the electron population
with $\tau=\Delta t_{inj}$ (i.e. the
oldest one); from Eq.(\ref{gamma(t)}),
with $\gamma(t_i)=\gamma_b^{rg}$, one has :

\begin{eqnarray}
\gamma_{max}=
{{ \gamma_b(\Delta t_{inj})
-\xi/(\beta \gamma_b^{rg}) }\over
{ 1 + [\gamma_b(\Delta t_{inj}) -
\chi /\beta ] (\gamma_b^{rg})^{-1}
}} 
\cases{
{ \buildrel \gamma_b^{rg} \rightarrow \infty  \over
\longrightarrow } \gamma_b(\Delta t_{inj})
\cr \cr
{ \buildrel \gamma_b^{rg} = \gamma_b(\infty)  \over = }
\gamma_b(\infty) 
\cr }
\label{gammamax}
\end{eqnarray}

\noindent
The integration is performed as follows (Fig A1):

\noindent
$\bullet \, \,$ case $\gamma_b^{rg} > \gamma_b(\infty)$ :

\begin{equation}
N(\gamma, \Delta t_{inj}) =
\int_0^{{\cal T}} N(\gamma, \tau) d\tau
\label{nfin}
\end{equation}

\noindent
where ${\cal T}= \Delta t_{inj}$ for
$\gamma < \gamma_{max}$, while 

\begin{equation}
{\cal T} = {1\over{\sqrt{q}}}
ln \left(
{{
\chi -\sqrt{q} 
-2\beta {\cal F}(\gamma)
}\over{
\chi +\sqrt{q} 
-2\beta {\cal F}(\gamma)
}} \right)
\label{tfin}
\end{equation}

\noindent
for $\gamma > \gamma_{max}$, where the function

\begin{equation}
{\cal F(\gamma)} = 
{{\gamma \gamma_b^{rg}}\over
{\gamma_b^{rg} - \gamma}}
\left\{
1+ S\left[\gamma; \gamma_b^{rg} \right]
\right\} 
\,
{ \buildrel \gamma_b^{rg} \rightarrow \infty  \over
\longrightarrow } \,
\gamma
\label{calf}
\end{equation}

\noindent
$\bullet \,\,$ case $\gamma_b^{rg} < \gamma_b(\infty)$ :

\begin{equation}
N(\gamma, \Delta t_{inj}) =
\int_{{\cal T}}^{ \Delta t_{inj} } 
N(\gamma, \tau) d\tau
\label{nfin2}
\end{equation}

\noindent
where ${\cal T}= 0$ for
$\gamma < \gamma_b^{rg}$, while it 
is given by Eq.(\ref{tfin})
for $\gamma > \gamma_b^{rg}$.

In the calculations we have not considered the
effect of statistical Fermi acceleration; this would
further complicate the equations.
In this case the main effect of 
the electron diffusion 
in the momentum space is a smoothing of the 
calculated features of the energy distribution.
In particular the energy cut off at $\gamma_{max}$
would become an exponential decrease.
However, the  
synchrotron and IC emitted spectrum are
not sensitive to minor differences in the
energy distribution being  
given by the convolution with 
the synchrotron and IC Kernel functions,  
respectively.
As a consequence, the systematic Fermi
approach is adequate to the aim of 
this paper.

\section{Appendix B: 
Synchrotron and IC formulae used in the
model computations}

\begin{figure}
\includegraphics{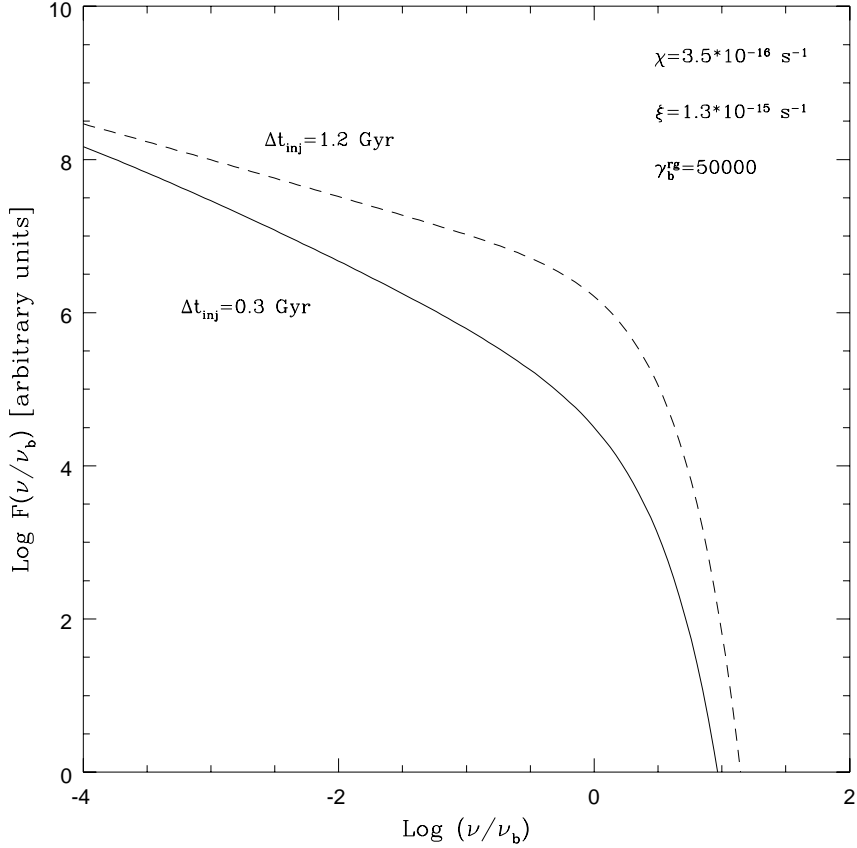}
\vspace{12 cm}
\caption{The calculated 
synchrotron spectrum  
from electrons continuously reaccelerated and 
injected in the ICM medium is reported for
different injection period $\Delta t_{inj}$ as
given in the panel.
The calculation is performed for  
$\chi=3.5 \cdot 10^{-16}$s$^{-1}$, 
$\xi=1.3 \cdot 10^{-15}$s$^{-1}$, 
$\gamma_b^{rg}=50000$ and
$\delta=2.6$.} 
\end{figure}

\noindent
In order to calculate the synchrotron emission
by the {\it AeP} we have assumed a 
magnetic field smoothly varying in space  
and tangled on a scale significantly
smaller than the 
variation scale of its intensity.
This allows us to consider the 
emitted synchrotron spectrum isotropic. 
In this case 
the synchrotron emissivity is obtained 
by integrating
the isotropic synchrotron Kernel
over the electron energy and angular
distribution 
(cf. Westfold 1959; Pacholczyk 1970).
The synchrotron emissivity per unit 
solid angle is:

\begin{eqnarray}
j^{S}\left( {{\nu}\over{\nu_b}}, 
\Delta t_{inj} \right)=
{{ \sqrt{3} }\over{8 \pi}}
{{e^3}\over{m c^2}}
{K_e B \over {  w^{\delta-1} }} { q \over{\beta^2}}
\int_0^{ {{\pi}\over 2} }
d \theta 
sin^2 \theta  
\nonumber\\
\int_0^1 du
F \left( {{ \nu/\nu_b }\over{
u^2 sin\theta }} \right) u^{-\delta}
\int_0^{ {\cal T}(u) } d\tau
\left\{
1+
S \left(u, \tau \right) 
\right\}^{-\delta}
\nonumber\\
\left(
1- u {{ w}\over{\gamma_b^{rg} }}
{{
1 + 
S\left(
u,\tau
\right) }\over
{
1- u w /\gamma_b(\tau) }}
\right)^{\delta-2} \cdot
\nonumber\\
\gamma_b^{-2}(\tau)
\left( 
1 - {1\over{tanh(x) }} \right)^2 
\left( 1- u {{w}\over{\gamma_b(\tau)}} \right)^{
\delta-2}
\label{syn}
\end{eqnarray}

\noindent
where 

\begin{equation}
S\left( u, \tau \right)
=
{{
\xi u^{-1} w^{-1} -\chi }\over
{ \beta \gamma_b(\tau) }}
\label{ss}
\end{equation}

\noindent 
$u = \gamma/w$, $w= \gamma_b^{rg}$ or $\gamma_b(\Delta t_{inj})$
if $\gamma_b^{rg} > \gamma_b(\infty)$ or 
$\gamma_b^{rg} < \gamma_b(\infty)$, respectively, 
$F$ is the isotropic 
electron Kernel (e.g. Pacholczyk 1970) and
the other quantities are defined in 
Appendix A.

\noindent
Because of the angular distribution of the
CMB photons is isotropic, 
the IC spectrum is obtained by integrating the 
electron energy distribution of the {\it AeP} 
over the 
isotropic IC Kernel in the ultrarelativistic case
(e.g. Blumenthal \& Gould 1970).
The assumptions for the magnetic field
distribution 
adopted in the case of
the synchrotron spectrum, guarantee the isotropy
of the IC emission even if the effect of 
synchrotron losses are important for 
the electron energy distribution.
The IC emissivity per unit solid angle is given 
by:

\begin{eqnarray}
j^{IC}\left( \epsilon_1 , 
\Delta t_{inj} \right)=
K_e 
{{ 2^{\delta+3} r_0^2 \pi }
\over{c^2 h^2 
}} {q \over {\beta^2}} 
\int {{ (\epsilon/\epsilon_1)^{ {{\delta+5}\over 2} } d\epsilon}\over{ 
exp\left\{\epsilon/kT_{cmb} \right\} - 1 }}
\nonumber\\
\int_{1}^{ y_{max} }
dy  y^{ {{\delta+1}\over2} } 
\left( 1 - 2{ y} +2 ln{ y} + 
{1\over { { y} }} \right)
\nonumber\\ 
\int_0^{ {\cal T}(\gamma) } d\tau
\left\{
1+ S\left( { y}, \tau \right)
\right\}^{-\delta}
\nonumber\\
\left\{
\left( 1-  {{ { y} }\over{ { y}_b(\tau)}} \right)
\left(
1-  {{ { y} }\over{ { y}_b^{rg} }}
{{
1 + 
S\left( { y}, \tau \right)
}\over
{
1- { y}/{ y}_b(\tau) }}
\right) \right\}^{\delta-2} 
\nonumber\\
y_b(\tau)
\left( 
1 - {1\over{tanh(x) }} \right)^2 
\label{ic}
\end{eqnarray}

\noindent
where 

\begin{equation}
S\left( y, \tau \right)
=
{{
\xi y^{-1} \left(y_b^{rg}\right)^{-1} -\chi }\over
{ \beta y_b(\tau) }}
\label{ss}
\end{equation}

\noindent
while, 
$y_b^{rg}= (\epsilon_1 / 4\epsilon)(\gamma_b^{rg})^{-2}$, 
$y_{max}= (\epsilon_1 / 4\epsilon)w^{-2}$, and 
$y_b(\tau)= (\epsilon_1 / 4\epsilon)(\gamma_b(\tau))^{-2}$.

\noindent
An example of the synchrotron 
spectrum obtained from Eq.(\ref{syn}) is given 
in Fig.B1.
Here, the break frequency is defined as the 
critical synchrotron frequency 
of the electrons with the maximum Lorentz 
factor (in this case $\gamma_b^{rg}$) and 
pitch angle $\theta$=90$^o$.
Following the evolution of the electron
energy distribution (e.g. Figs. 1 \& 2)
the synchrotron spectrum flattens with increasing
$\Delta t_{inj}$.

\vskip 1 truecm

{\bf References}

\par\medskip\noindent
Arabadjis J.S., Bregman J.N., 1999, Apj 514, 607
\par\medskip\noindent
Atoyan A.M., V\"olk H.J., 2000, in press;
astro-ph/9912557
\par\medskip\noindent
Bergh\"ofer T.W., Bowyer S., Korpela E., 2000, ApJ 535,
615
\par\medskip\noindent
Blumenthal G.R., Gould R.J., 1970 Rev. of Mod. Phys. 
42, 237
\par\medskip\noindent
Bonamente M., Lieu R., Mittaz J.P.D., 2000, ApJ L in press; 
astro-ph/0011186
\par\medskip\noindent
Bowyer S., Lampton M., Lieu R., 1996, Science 274, 
1338
\par\medskip\noindent
Bowyer S., Bergh\"{o}fer T.W., 1998, ApJ 506, 502
\par\medskip\noindent
Bowyer S., Bergh\"{o}fer T.W., Korpela E., 1999, ApJ
526, 592
\par\medskip\noindent
Brunetti G., Setti G., Comastri A., 1997, 
A\&A 325, 898
\par\medskip\noindent
Brunetti G., Feretti L., Giovannini G., Setti G., 1999
in {\it Diffuse Thermal and Relativistic Plasma in Galaxy
Clusters}, eds. H.B\"{o}hringer, L.Feretti, P.Schuecker, 
MPE Report 271, p.263 
\par\medskip\noindent
Brunetti G., Setti G., Feretti L., Giovannini G., 
2000, MNRAS in press; astro-ph/0008518
\par\medskip\noindent
Donnelly R.H., Markevitch M., Forman W., 
Jones C., Churazov E., Gilfanov M., 1999, ApJ 513, 690
\par\medskip\noindent
Ensslin T.A., Biermann P.L., Kronberg P.P.,
Wu X.-P., 1997, ApJ 477, 560
\par\medskip\noindent
Ensslin T.A., Lieu R., Biermann P.L., 1999, A\&A
344, 409
1978, A\&A 69, 253
\par\medskip\noindent
Dallacasa D., Feretti L., Giovannini G., Venturi T., 
1989, A\&AS 79, 391
\par\medskip\noindent
Feretti L., Dallacasa D., Giovannini G., Venturi T., 
1990, A\&A 232, 337
\par\medskip\noindent
Feretti L., Dallacasa D., Giovannini G., Tagliani A., 
1995, A\&A 302, 680
\par\medskip\noindent
Feretti L., Giovannini G., 1996, IAUS 175, 347
\par\medskip\noindent
Feretti L., Giovannini G., Klein U., et al., 1998, 
A\&A 331, 475
\par\medskip\noindent
Fusco--Femiano R., Dal Fiume D., Feretti L., et al., 1999
ApJ 513, 197L
\par\medskip\noindent
Giovannini G., Feretti L., Venturi T., 
Kim K.T., Kronberg P.P., 1993, ApJ 406, 399
\par\medskip\noindent
Giovannini G., Tordi M., Feretti L., 
1999, NewA 4, 141. 
\par\medskip\noindent
Kardashev N.S., 1962, Sov. Ast. 6, 317
\par\medskip\noindent
Lieu R., Mittaz J.P.D., Bowyer S., et al., 1996,
Science 274, 1335
\par\medskip\noindent
Lieu R., Mittaz J.P.D., Bowyer S., et al., 1996,
ApJ 458, L5
\par\medskip\noindent
Lieu R., Ip W.-H., Axford W.I., Bonamente M., 1999, 
ApJ 510, L25
\par\medskip\noindent
Lieu R., Bonamente M., Mittaz J.P.D., 2000, A\&A, in press;
astro-ph/0010610
\par\medskip\noindent
Murgia M, Fanti R., 1996, IRA {\it internal report}, 228/96
\par\medskip\noindent
Pacholczyk A.G., 1970, Radio Astrophysics, eds. G. Burbidge
\& M. Burbidge, Freeman and Company, San Francisco
\par\medskip\noindent
Sarazin C.L., 1999, ApJ 520, 529
\par\medskip\noindent
Sarazin C.L., 2000, in {\it Large Scale Structure in
the X--ray Universe}, p.81, eds. M.Plionis, I. Georgantopoulos
\par\medskip\noindent
Sarazin C.L., Lieu R., 1998, ApJL 494, 177
\par\medskip\noindent
Saripalli L., Subrahmanyan R., Hunstead R.W., 1994, 
MNRAS 269, 37
\par\medskip\noindent
Schoenmakers A.P., Mack K.-H., Lara L., et al., 1998, 
A\&A 336, 455
\par\medskip\noindent
V\"{o}lk H.J., Atoyan A.M., 1999, APh 11, 73

\end{document}